\documentclass[a4paper,11pt]{article}
\pdfoutput=1 

\usepackage{jinstpub} 

\usepackage{lineno,hyperref}
\usepackage{lipsum}
\usepackage{graphicx}
\usepackage{numcompress}
\usepackage{multirow}
\modulolinenumbers[5]
\usepackage{xcolor}
\usepackage{comment}
\usepackage{lineno}
\usepackage{python}
\usepackage{float}

\title{Machine Learning for Imaging Cherenkov Detectors}

\newcommand{\GlueX}{\textsc{GlueX}}

\author[a,b,1]{C. Fanelli\note{Corresponding author.}}

\affiliation[a]{Laboratory for Nuclear Science, Massachusetts Institute of Technology, Cambridge, MA 02139, USA}
\affiliation[b]{Jefferson Lab, EIC Center, Newport News, VA 23606, USA}

\emailAdd{cfanelli@mit.edu}

\abstract{

Imaging Cherenkov detectors are largely used in modern nuclear and particle physics experiments where cutting-edge solutions are needed to face always more growing computing demands.

This is a fertile ground for AI-based approaches and at present we are witnessing the onset of new highly efficient and fast applications.  
This paper focuses on novel directions with applications to Cherenkov detectors. 
In particular, recent advances on detector design and calibration, as well as particle identification are presented.
}


\proceeding{International Workshop on Fast Cherenkov Detectors - \\Photon detection, DIRC design and DAQ,\\
  11 - 13 September 2019, \\
  Castle Rauischholzhausen, Justus-Liebig-University Giessen, Germany}

\begin{document}
\maketitle
\flushbottom


\section{Introduction}\label{sec:intro}

Cherenkov detectors are largely used in modern nuclear and particle physics experiments for charged particle identification (PID). Cherenkov radiation is released in the shape of a cone along the particle momentum direction when a charged particle moves through a dielectric medium at a speed larger than the phase velocity of light in that medium.
A Cherenkov detector is typically equipped with single photon detectors, and the charged particles can be identified by the shapes of the detected hit patterns.

The plenitude of photon detectors gives rise to a pattern recognition
task being the original master domain of Artificial Intelligence (AI).
AI is becoming ubiquitous in our field, particularly in high energy physics \cite{albertsson2018machine}: following a standard taxonomy \cite{MEHTA20191},
generally AI encompasses all the concepts related to the integration of human intelligence into machines; a subset of AI is machine learning (ML), which means empowering machines to learn, and typically these algorithms can be grouped into supervised, unsupervised and reinforcement learning; whereas deep learning (DL), a subset of ML---often considered the evolution of ML---is based on deep (\textit{i.e.}, made by many hidden layers) neural networks. 
In the most frequent applications, features are selected and a model is trained for classification or regression using signal and background examples. 

Artificial intelligence can have an impact on a wide range of activities in our field, \textit{e.g.}, DAQ, monitoring, calibration, data reconstruction and analysis, detector design etc. 

This proceeding focuses on existing approaches and novel directions with specific applications to imaging Cherenkov detectors.

The outline of this paper is as follows: existing reconstruction methods including different areas of research are discussed in Sec. \ref{sec:sota}; novel directions (detector calibration, design optimization, and PID) are presented in Sec. \ref{sec:novel}; summary and conclusions are reported in Sec. \ref{sec:summary}.

\section{State of the Art}\label{sec:sota}

First applications of Artificial Neural Networks (ANN) \cite{hopfield1988artificial} for pattern recognition in ring-imaging Cherenkov (RICH) detectors date back to the nineties \cite{castellano1991artificial}, where a two-layer neural network with forward connections was proposed to evaluate the analysis of a set of images produced by an optical RICH detector at CERN.

More recent applications of ANN to pattern recognition have been realized for the Compressed Baryonic Matter (CBM) experiment \cite{lebedev2009fast}, which aims to measure dileptons emitted from the hot and dense phase in heavy-ion collisions. Electron identification is performed by a RICH and transition radiation detectors. 
Data processing plays an important role, and in this case fast and efficient algorithms based on fast Hough transform are used for the ring search. 
Ultimately, ANN is used for ring classification. In particular, a detailed study of the procedure of fake ring elimination is performed taking into account the optical distortions resulting in ellipse fitting methods and radius corrections. 

Another interesting application comes from the LHCb experiment \cite{lhcb2013lhcb}, where ML-based global PID algorithms are developed to improve the identification of particles combining the information associated to several sub-detectors \cite{derkach2019machine}: two RICH sub-detectors provide charged hadrons ($\pi$,$K$,$p$) identification over a wide momentum range, from 2 to 100 GeV/c. Muons ($\mu$) are identified mainly by dedicated muon chambers, while electron ($e$) and photon ($\gamma$) identification is assured by the calorimeters. 
Global PID can be obtained by simply computing Log Likelihoods (LL) separately for each sub-detector and then combining them. It has been proved though that more refined methods like ANN outperform the LL-based approach. 
Furthermore, dedicated methods (see, \textit{e.g.}, \cite{rogozhnikov2015new}) have been developed to improve the flatness of the reconstruction efficiencies as a function of the particle phase-space in order to reduce systematic uncertainties. 

The Very Energetic Radiation Imaging Telescope Array System (VERITAS) \cite{holder2006first} has been one of the first experiments to make use of DL for detection of Cherenkov rings.
The Muon Hunter \cite{bird2018muon} is an interesting project hosted on the Zooniverse platform, where volunteers select pictures of data from the VERITAS cameras to identify muon ring images.
Volunteering work is the basis to obtain a reliable set of training data to finally mimic their human experience by ML. 

VERITAS is made by four 12m diamater imaging atmospheric Cherenkov telescopes sited at the Fred Lawrence Whipple Observatory in southern Arizona.
The identification of muon ring images is based on a Convolutional Neural Network (CNN) \cite{lecun1995convolutional} which is capable of rejecting background events and identifying suitable calibration data to monitor the telescope performance as a function of time.
The supervision of the volunteers provided a more efficient machine learning model and helped identifying unexpected images. 

In neutrino experiments, water Cherenkov detectors are commonly used to distinguish between $e$/$\mu$ leptons, which determine the flavour of the interacting neutrino in the medium. 
An example of this kind of detectors is Super-Kamiokande \cite{fukuda2003super}, where the use of ML always in form of CNN based on Tensorflow \cite{abadi2016tensorflow} has been explored \cite{theodore2016particle}. Notice that the standard analysis fitting algorithm is specifically tuned for the problem of $e$/$\mu$ separation and is based on complex modelling of Cherenkov light moving and scattering in the detector. The DL approach, on the other hand, has the advantage that it requires no prior knowledge of the Cherenkov physics, and all these features are directly learned by the algorithm which turns out to have performance comparable to the the standard PID.
Application of ML techniques to reconstruct lepton energies in water Cherenkov detectors is proposed for TITUS \cite{lasorak2015titus}---an intermediate detector for the Hyper-Kamiokande experiment. It has been found in \cite{drakopoulou2018application} that this leads to more than 50\% improvement in the energy resolution for all lepton energies compared to standard approaches based upon lookup tables, with performance comparable to likelihood-function based techniques currently in use.

\section{Novel Directions}\label{sec:novel}

This section describes novel AI-based methods applied to imaging Cherenkov detectors in nuclear and particle physics: methods for detector design and calibration based on Bayesian optimization are described in Sec. \ref{sec:bayes}; PID applications based on DL are described in Sec. \ref{sec:DL}.

\subsection{Bayesian Optimization}\label{sec:bayes}

Bayesian Optimization (BO) \cite{jones1998efficient, snoek2012practical} is among the most efficient tools for tuning the parameters of a black-box functions $f(x)$, searching for the global optimum $x^{*}$ over a bounded domain $\chi$ of $f$.
In particular, $f$ can be noisy, non-differentiable and expensive to evaluate. 
Typically Gaussian processes \cite{williams2006gaussian} are used to build a surrogate model of $f$, but other regression methods such as decision trees can also be used. 
Once the probabilistic model is determined, a cheap utility function (also called acquisition function) is considered to guide the process of sampling the next point to evaluate. 

In the following, two applications are described: in Sec.~\ref{sec:bayes-calib} BO is applied to detector calibration and in Sec.~\ref{sec:bayes-design} to detector design.

%
%

\subsubsection{Detector Calibration}\label{sec:bayes-calib}

This section shows how Bayesian optimization can be applied to detector calibration. 
As an example, we consider the DIRC (acronym of Detection of Internally Reflected Cherenkov light) detector recently installed in \GlueX \ \cite{gluex2016first}, a particle physics experiment located in Hall D at the Jefferson Lab (JLab) \cite{dudek2012physics}. 

\begin{figure}[htbp]
\centering 
\includegraphics[width=.5\textwidth,origin=c,angle=0]{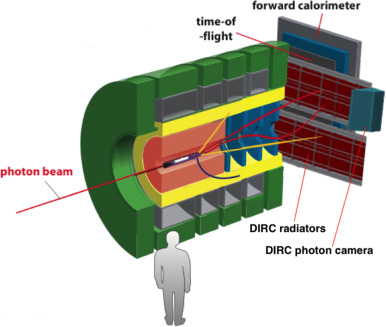}
\qquad
\includegraphics[width=.35\textwidth,origin=c,angle=0]{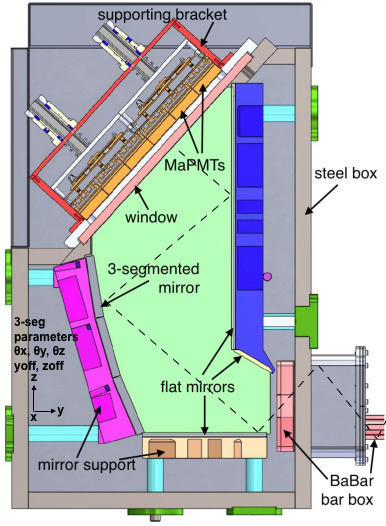}
\caption{\label{fig:GlueX-DIRC} (left)  A schematic view of the \GlueX \ detector with the DIRC, which consists of four radiator boxes and two photon cameras; (right) Schematic diagram of the cross-section of the photon camera. The attached bar boxes are shown on the bottom right. The bottom flat mirror is aligned with the radiators. Image taken from~\cite{barbosa2017gluex}.}
\end{figure}

The DIRC will improve the \GlueX \ PID capabilities in the forward region (see Fig. \ref{fig:GlueX-DIRC} (left)), and it is essential for the physics program of \GlueX, whose primary goal is to search for and ultimately study the properties of hybrid mesons \cite{meyer2015hybrid}.
It consists of four bar boxes oriented to form a plane 4~m away from the fixed target of the experiment and two photon cameras. 
Each bar box contains 12 fused silica radiators (1.725$\times$3.5$\times$490 cm$^{3}$) with a small wedge attached to the end which is read out.
The DIRC photon camera contains multiple flat mirrors to direct the light to the photodetector plane. 
Each photon camera is attached to two bar boxes and is equipped with an array of $\sim$100 Hamamatsu H12700 MaPMTs \cite{calvi2015characterization}. 
A DIRC employs rectangular fused-silica bars both as Cherenkov radiators and as light guides. 
Typically the detected hit pattern in the PMT plane is sparse making the reconstruction rather challenging.  
 The readout electronics boards are the same as for the CLAS12 RICH \cite{el2012rich} in Hall B at JLab. 

During data taking, when the optical box is filled with distilled water, several components could be misaligned. 
Consequently, a set of ($\gtrsim$10) main alignment parameters should be monitored (see Fig. \ref{fig:GlueX-DIRC} (right)), \textit{e.g.}, the three-segmented mirror angles and spatial displacement, the position of the bars relative to the outside of the bar box, the relative distance of the mirror to PMT plane, etc. 
For all these reasons, the DIRC calibration can be considered a black-box problem---with potentially many non-differentiable terms---in a noisy environment.

In the following the vector of offsets is called $\vec{\theta}$, and   results shown are taken from a preliminary closure test presented in \cite{bayesian_DIRC:2018} (where the reader can find more details about the estimated tolerances).

The calibration strategy proposed in \cite{bayesian_DIRC:2018} allows to \textit{self-learn} the offsets from observations of real data. 
The idea is to use a large sample of pions with high-purity obtained from channels like $\rho$ decays which are abundantly produced in photoproduction experiments like \GlueX. 
At low momentum these charged particles are well identified by other subdetectors in \GlueX \ (without the DIRC contribution) and can be used as high-purity sample for the calibration procedure. 

The objective function can be any figure of merit (FoM) (in the form of, \textit{e.g.}, a log-likelihood) that compares noisy observations---the sparse hit pattern detected in the PMT plane---with the expected hit pattern under certain hypotheses: the particle ID, momentum, location and orientation of the charged particle traversing the bars, and the offsets. 
These preliminary studies are based on FastDIRC \cite{hardin2016fastdirc}, a fast Monte Carlo and reconstruction algorithm for DIRC detectors which is more than 10$^{4}$ times faster than \textup{Geant}-based simulation and allows to parameterize the effect of the offsets.
About $\mathcal{O}(10^{4})$ charged particles are selected to form a high-purity sample by imposing certain requirements, like fiducial cuts on the incident and azimuthal angles with respect to the normal to the plane containing the fused silica bars and a minimum distance from the bar center.

The closure test consists basically in injecting a known set of offsets and reverse-engineer them within the allowed tolerances.
This has been proven with 7 main parameters and has been tested for different values of the offsets.
 A framework based on BO has been used, which allows to perform the calibration in a much reduced amount of time compared to a simple random search (RS). Time performance will be further optimized in future studies. 

The closure test is completed by using the determined offsets to calibrate the data, and check if the detector performance is consistent with what one would obtain with the true offsets. 
Good proxies to evaluate the quality of the reconstruction are the single photon resolution (SPR), or the effective resolution (SPR normalized to the root of the photon yield), and the area under curve (AUC) of a ROC curve \cite{hanley1982meaning}. 
These quantities can be evaluated considering $\pi$s and $K$s at different kinematics. 

\begin{figure}[htbp]
\centering 
\includegraphics[width=.32\textwidth,origin=c,angle=0]{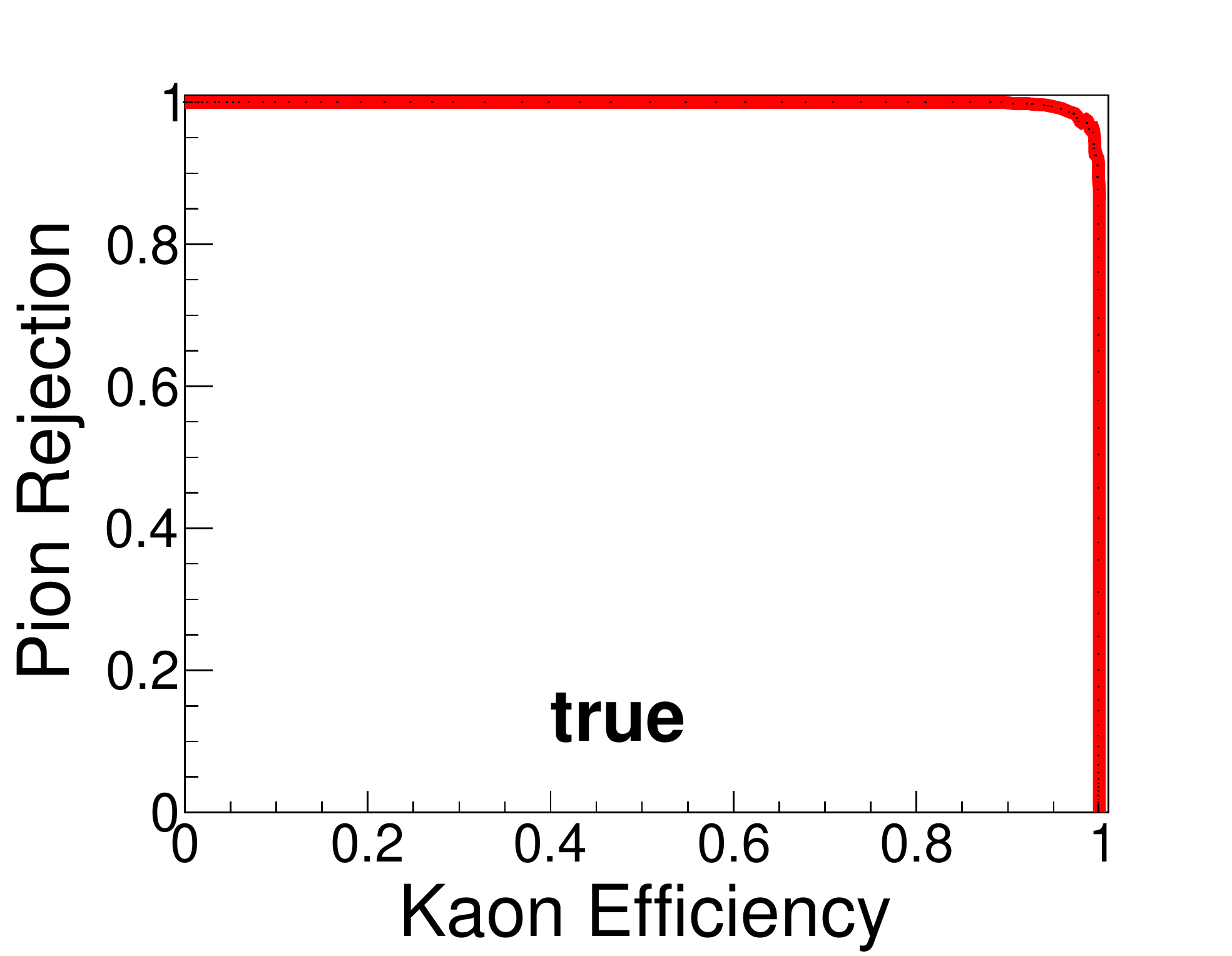}
\includegraphics[width=.32\textwidth,origin=c,angle=0]{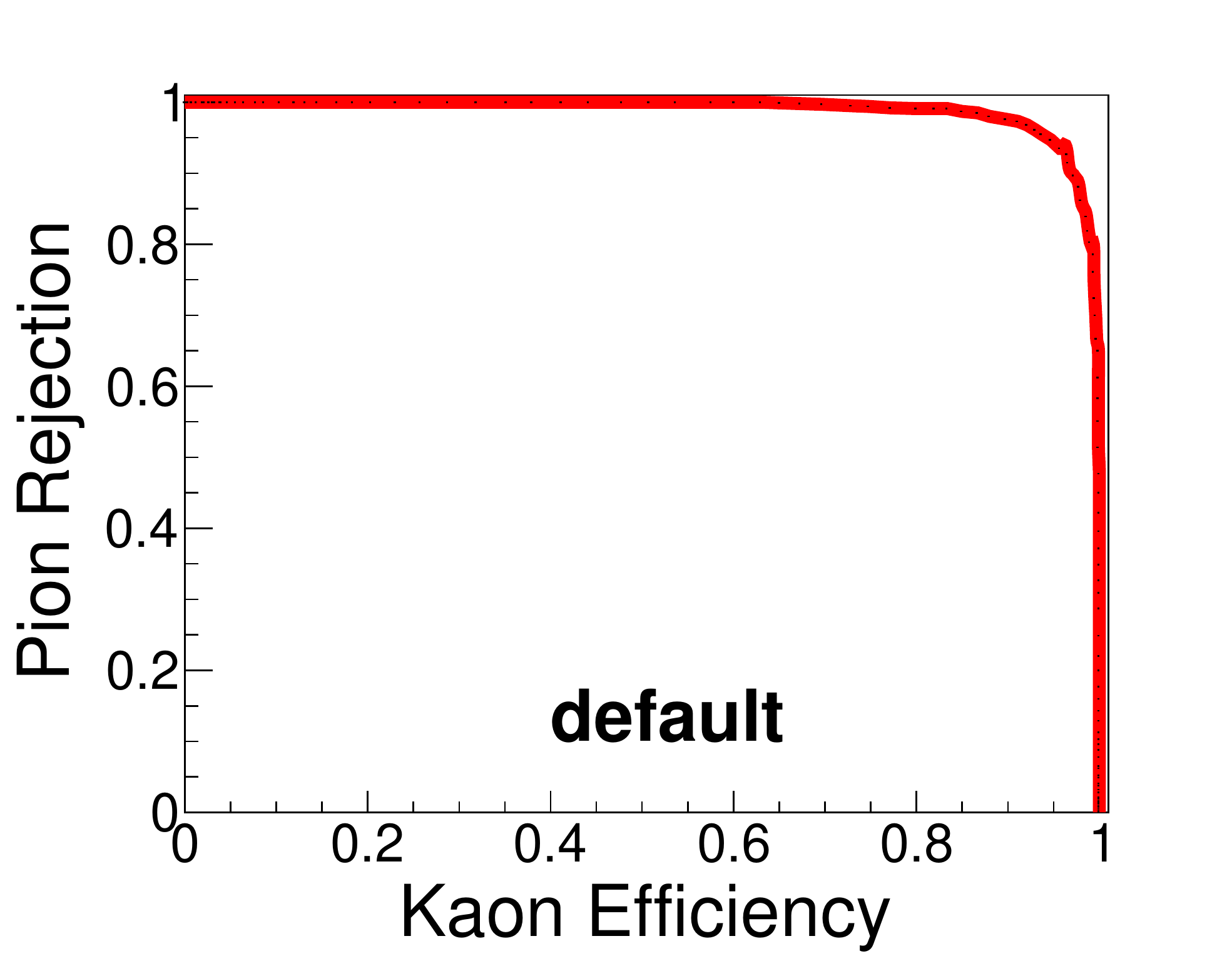}
\includegraphics[width=.32\textwidth,origin=c,angle=0]{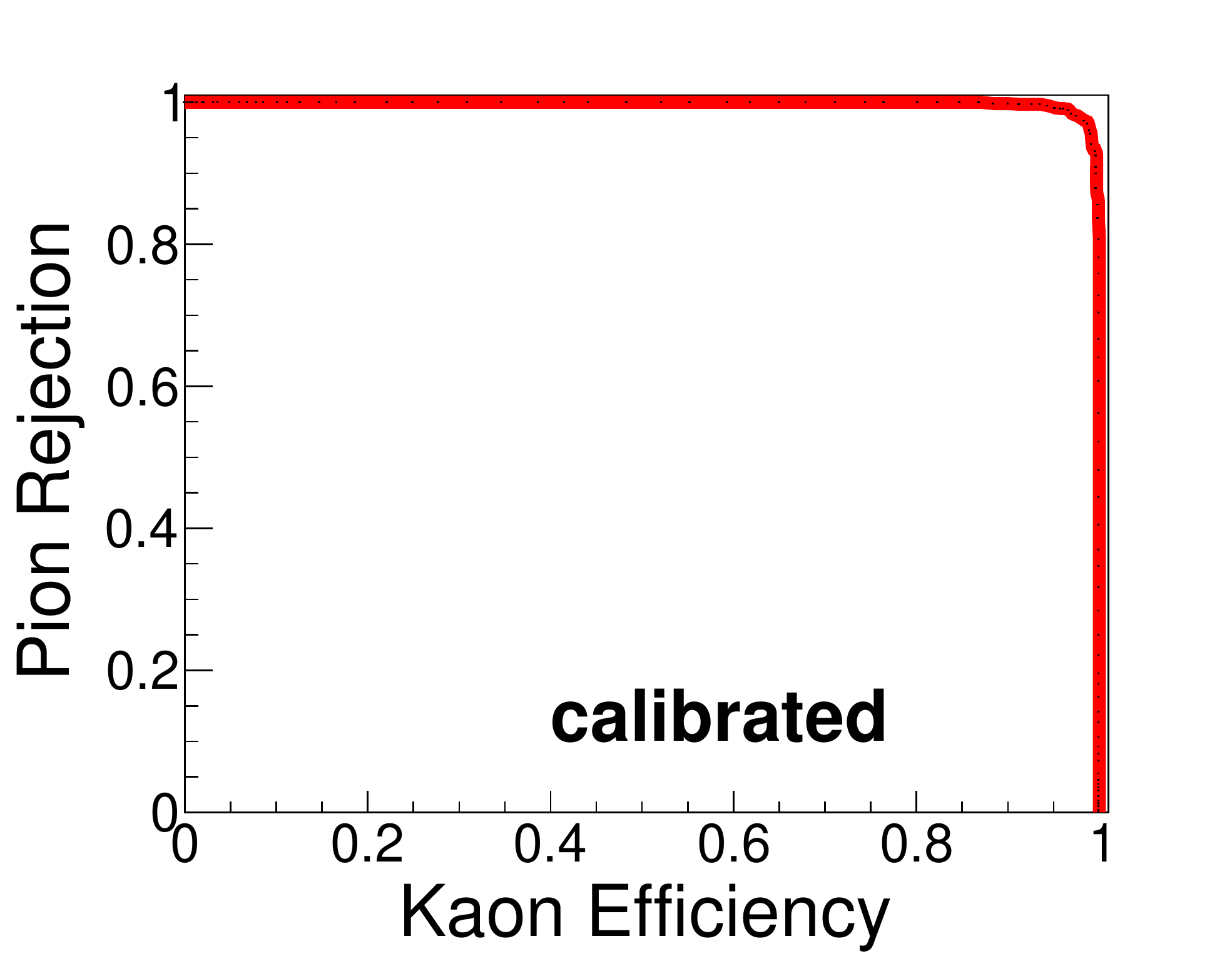}
\caption{\label{fig:ROCcurves} ROC curves (pion rejection vs kaon efficiency) for particles with kinematics (P,$\theta$,$\phi$): (4 GeV, 4 deg, 40 deg). From left to right, (i) using the `true' offsets, (ii) without calibration (offsets set to default null values), (iii) after calibration based on the results of the BO.  The AUC is used as proxy for the closure test. Further details can be found in \cite{bayesian_DIRC:2018}.
}
\end{figure}

Examples of ROC curves obtained in the closure test are shown in Fig. \ref{fig:ROCcurves}: (i) the left plot corresponds to the true values of the offsets; (ii) the middle plot is without calibration; (iii) the right plot shows the ROC after calibration.  
As shown in \cite{bayesian_DIRC:2018}, the values of the SPR and the AUC obtained after the calibration are consistent with the expected ones, proving the quality of the calibration procedure. 

\subsubsection{Detector Design}\label{sec:bayes-design}

This section shows an example of AI applied to detector design and is based on a recent work done by \cite{cisbani2019ai}, where a highly parallelized and automated procedure using Bayesian optimization and machine learning that encodes detector requirements is proposed.  

\begin{figure}[!ht]
\centering
\includegraphics[scale=0.5, angle = 0]{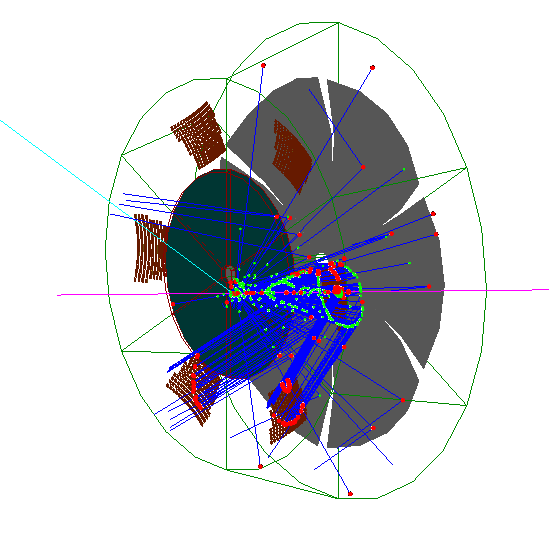}
\includegraphics[scale=0.45, angle = 0]{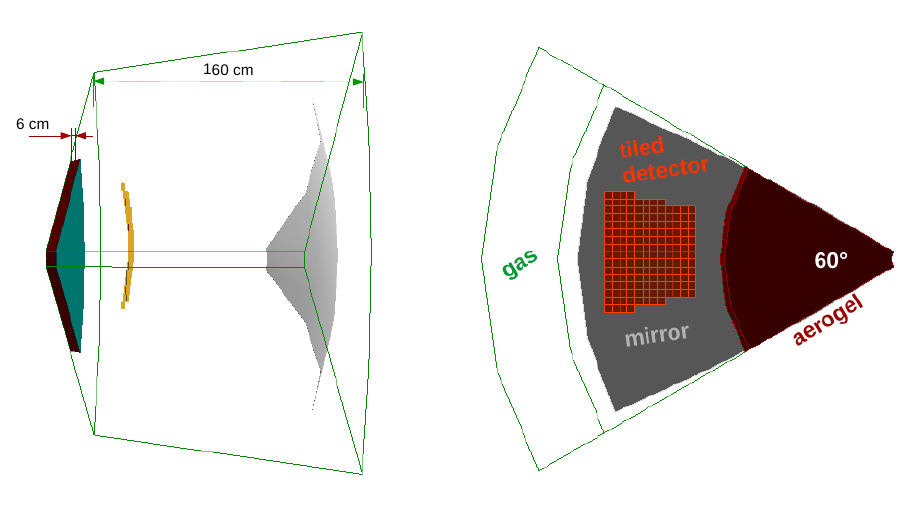}
\caption{Geant4 based simulation of the dRICH. In transparent wired red is the aerogel radiator, in transparent wired green is the gas radiator volume; the mirrors sectors are in gray and the photo-detector surfaces (spherical shape) of about 8500~cm$^2$ per sector in dark-yellow. A pion of momentum 10~GeV/c is simulated. Image taken from \cite{cisbani2019ai}.}
\label{fig:dual1}
\end{figure}

The design of the dual-radiator Ring Imaging Cherenkov (dRICH) \cite{akopov2002hermes,lhcbdetector,adinolfi2013performance} detector---under development as part of the particle-identification system at the future Electron-Ion Collider (EIC) \cite{eichandbook_2019}---is considered as a case study. 
The baseline design consists of two radiators (aerogel and C$_2$F$_6$ gas) sharing the same outward-focusing spherical mirror and highly segmented ($\approx 3$\,mm$^2$ pixel size) photosensors located outside of the charged-particle acceptance. 
Details of the dRICH detector design are shown in Fig.~\ref{fig:dual1}.

\begin{figure}[t]
\centering
\includegraphics[scale=0.325, angle = 0]{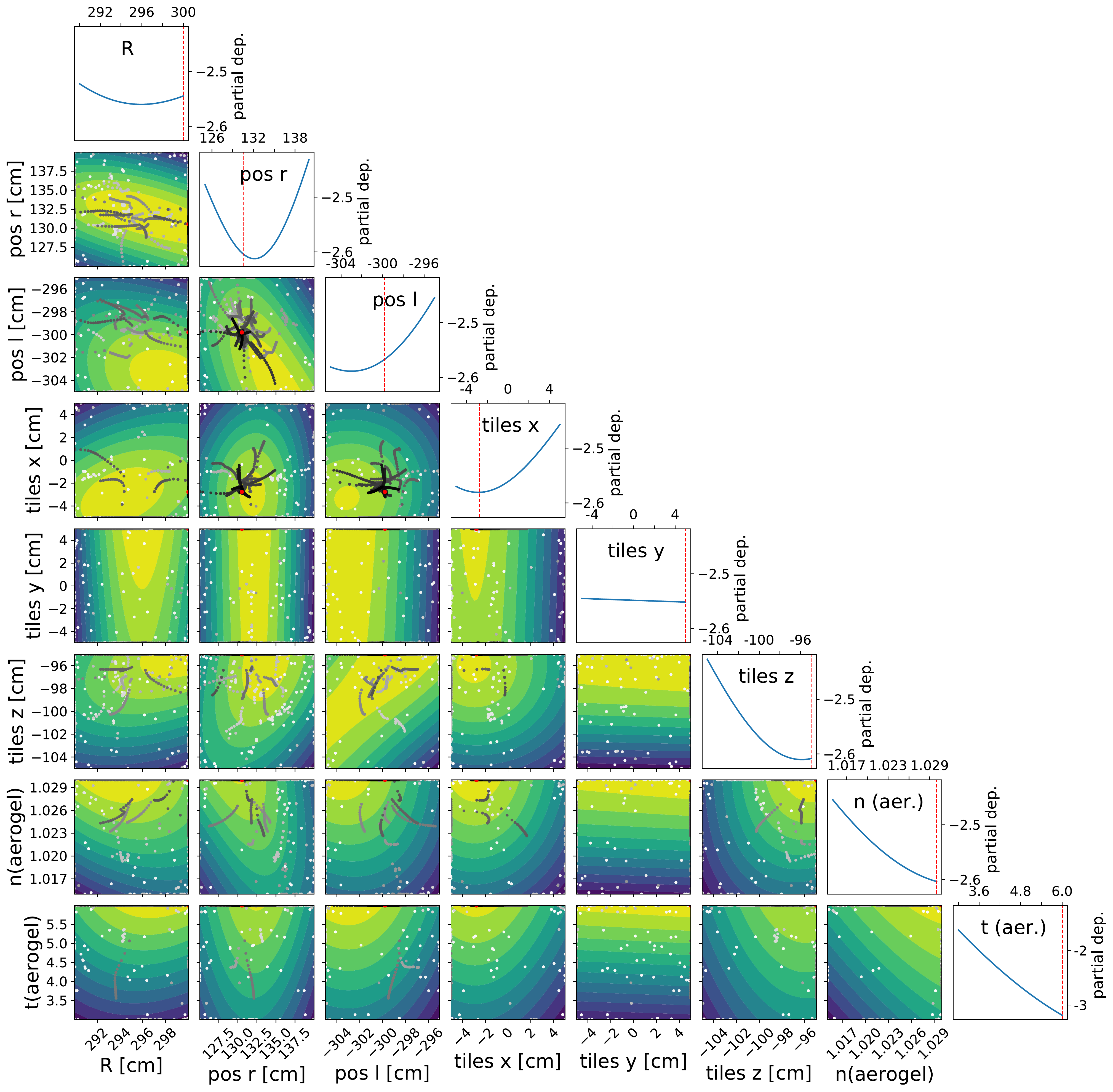}
\caption{2D plot of the objective function (color axis). The optimization strategy of the dRICH design involves tuning 8 parameters.  
In order to study possible correlations, each parameter is drawn against the other. The evaluations made by the optimizer are shown through an intensity gradient in the point trail ranging from white (first call of parallel observations) to black (last call). After about 55 calls, the stopping criteria are activated. The dotted red lines correspond to the projections on each variable of the optimal point found by the BO. 
More details can be found in \cite{cisbani2019ai}.
}
\label{fig:posterior}
\end{figure}

\begin{figure}[!]
\centering 
\includegraphics[width=.65\textwidth,origin=c,angle=0]{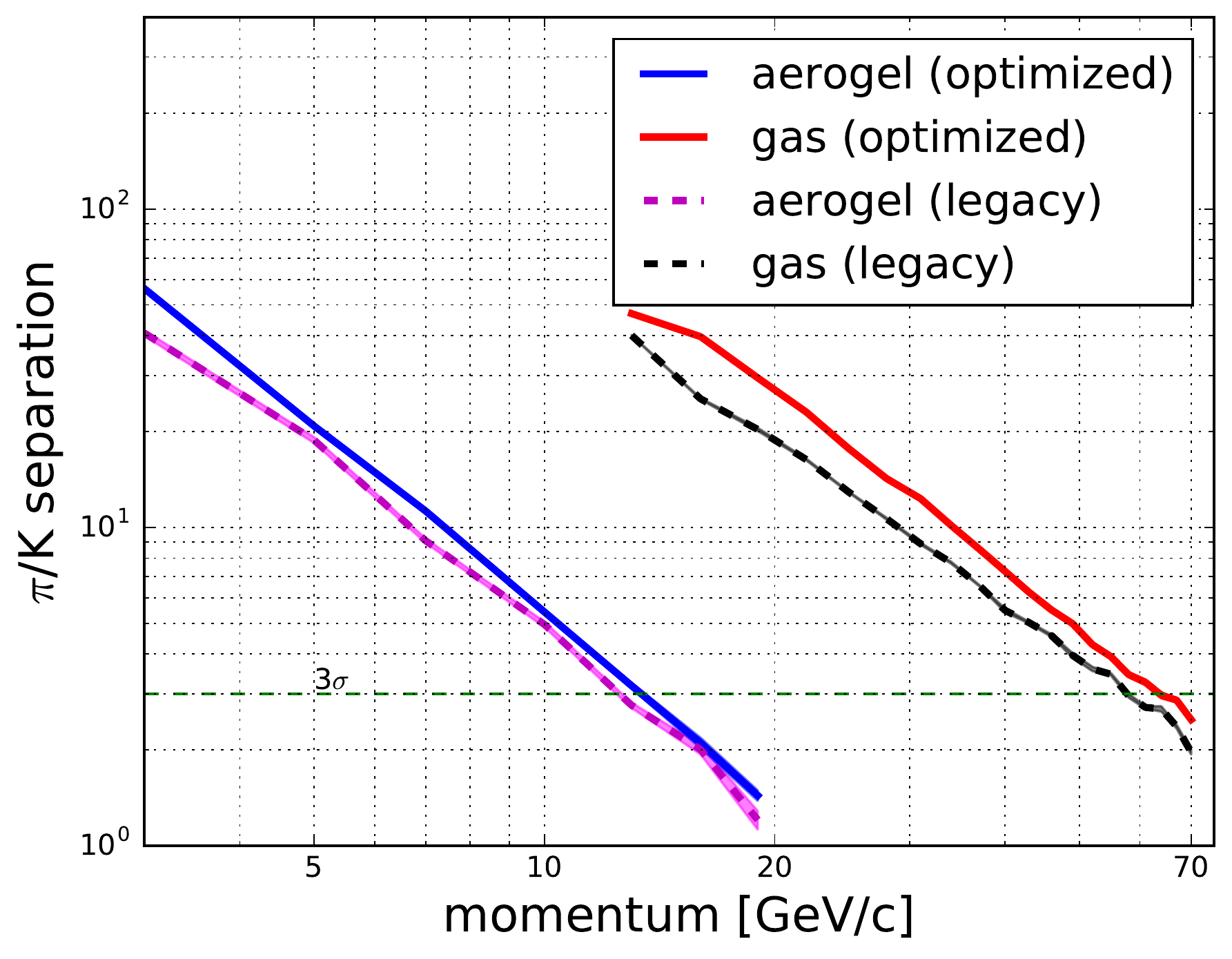}
\caption{\label{fig:improvement_baseline} 
$\pi/K$ separation as number of $\sigma$, as a function of the charged particle momentum. 
The plot shows the improvement in the separation power with the approach discussed in \cite{cisbani2019ai} compared to the legacy baseline design \cite{del2017design}. The curves are drawn with 68\% C.L. bands.  
}
\end{figure}

For the dRICH design optimization, eight parameters are considered to improve the PID performance: the refractive index and thickness of the aerogel radiator; the focusing mirror radius, its longitudinal (which determines the effective thickness of the gas) and radial positions (corresponding to the axis going in the radial direction in each of the six mirror sectors, see Fig.~\ref{fig:dual1}); and the 3D shifts of the photon sensor tiles with respect to the mirror center on a spherical surface, which to some extent determines the sensor area and orientation relative to the mirror.
These parameters, cover rather exhaustively the two major components of the dRICH: its radiators and optics. The regions of parameter space explored are based on previous studies~\cite{del2017design} under the constraint that it is possible to implement any values resulting from the optimization with (at worst) only minor hardware issues to solve.

Since the aim of the design optimization is to maximize the PID performance of the dRICH---to provide full hadron identification from a few GeV/c up to large momentum values---the separation power between pions and kaons has been as the objective function. 
In order to simultaneously optimize the combined PID performance of both the aerogel and gas parts in the dRICH, two working points have been determined based on the performance of the baseline design. 
Figure~\ref{fig:posterior} shows the posterior distribution in 2-dimensional subspaces of the design parameter space. These plots illustrate the possible correlations among the parameters. The optimal point in each subspace is marked with a red dot. Notice that the black points, corresponding to the points evaluated by the BO in its ask-and-tell procedure, tend to form basins of attraction around the minimum. 

Stopping criteria are developed to automatize the procedure which converges within a reduced number of iterations compared to RS. 
A detailed comparison between BO and RS is performed in \cite{cisbani2019ai} both in terms of PID and time performance.

The PID capabilities of the dRICH detector are substantially improved using the AI-based approach: the solid curves shown in Fig. \ref{fig:improvement_baseline} correspond to the new optimized design whereas the dashed curves are related to the previous baseline design \cite{del2017design}.
The developed procedure allows also to estimate the expected tolerances from the posterior distribution, within which any variation of the parameters does not alter the determined detector performance. 

Currently, there are many ongoing efforts to simulate and analyse EIC detector designs, and the same approach can be employed for any such study and can be extended to a global PID optimization of multiple detectors.
Interestingly, real-world costs of the components could be also included in the optimization method.

\subsection{Deep Learning}\label{sec:DL}

This section shows recent applications of deep learning applied to the DIRC detector for fast simulation and reconstruction of charged particles.    
Similar approaches can in principle be applied to other imaging Cherenkov detectors. 

\subsubsection{Fast Simulation and PID}\label{sec:DLgan}

A deep learning application to simulate Cherenkov detector response appeared recently in \cite{derkach2019cherenkov}, where it has been proposed to use a generative adversarial neural network (GAN) \cite{goodfellow2014generative} to bypass low-level details at the photon generation stage. 
This work is based on events simulated with FastDIRC \cite{hardin2016fastdirc} assuming the design of the \GlueX \ DIRC previously discussed in Sec.~\ref{sec:bayes-calib}.
\\The GAN architecture is trained to reproduce high-level features based on input observables of the incident charged particles, allowing for an improvement in simulation speed. 
The authors of this work claim a good precision and very fast performance from their studies. 

A novel deep learning algorithm for fast reconstruction has been proposed in \cite{fanelli2019deeprich} which can be applied to any imaging Cherenkov detectors. 
The core of this architecture is a generative model which leverages on a custom Variational Auto-encoder (VAE) \cite{VAE} combined to Maximum Mean Discrepancy (MMD) \cite{MMD}, with a Convolutional Neural Network (CNN) \cite{lecun1995convolutional} extracting features from the space of the latent variables for classification. 
A VAE is a particular type of generative models trying to simulate how the data are generated, in order to understand the underlying causal relations.

To this end, an Encoder produces a vector of latent variables by taking as input certain kinematic parameters---namely: the momentum $P$, the polar and azimuthal angles with respect to the normal to the radiator bars, $\theta$ and $\phi$, as well as the position where the charged particle is traversing the bar ($X$,$Y$)---concatenated with the hit pattern of the charged particle.
The vectors of latent variables associated to the hits of a particle are used to classify the particle. 
The Decoder reconstructs the input hits using the latent variables and the kinematic parameters.
Standard regularization techniques (\textit{e.g.}, dropout, batch normalization) are considered in order to prevent overfitting and improve performance and stability of the network (more details can be found in Table~1 of \cite{fanelli2019deeprich}).  

\begin{figure}[!]
\centering
\includegraphics[width=.6\textwidth]{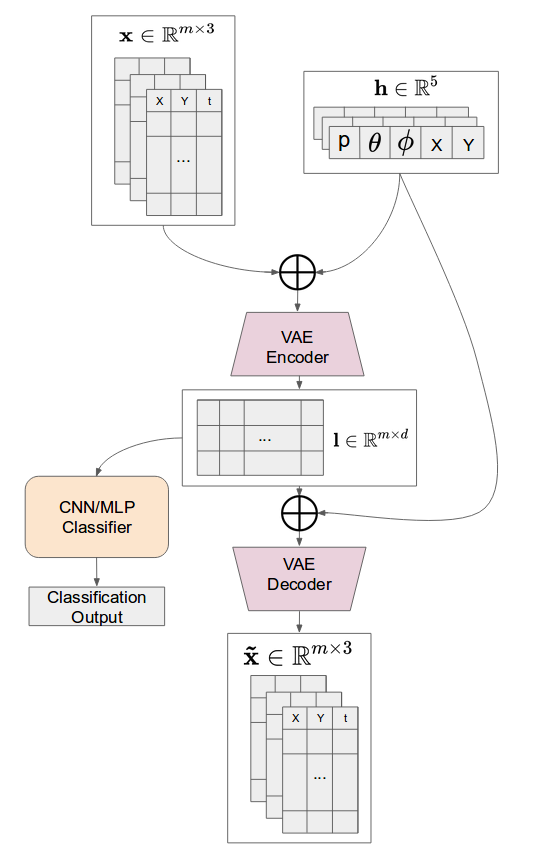}
\caption{A flowchart of DeepRICH: the inputs are concatenated---n.b., the $\oplus$ represents the concatenation between vectors---and injected into the encoder, which generates a set of vectors of latent variables, which are then used for both the classification of the particle and for the reconstruction of the hits.
Image taken from \cite{fanelli2019deeprich}.
}
\label{fig:model}
\end{figure}

A flowchart of the DeepRICH network is represented in Fig.~\ref{fig:model}.

\begin{figure}[H]
    \centering
    \includegraphics[width=.4\textwidth]{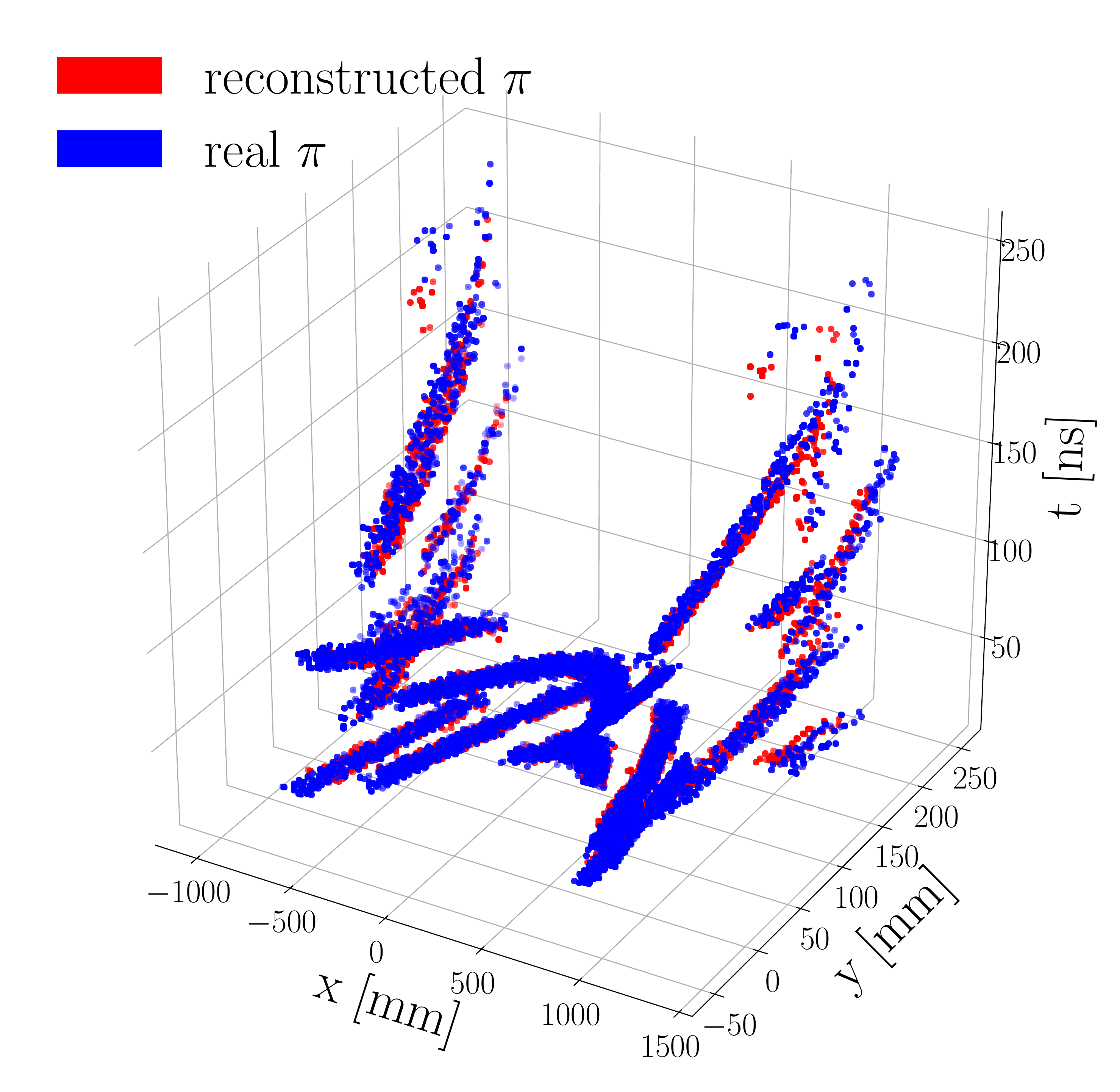}
    \includegraphics[width=.4\textwidth]{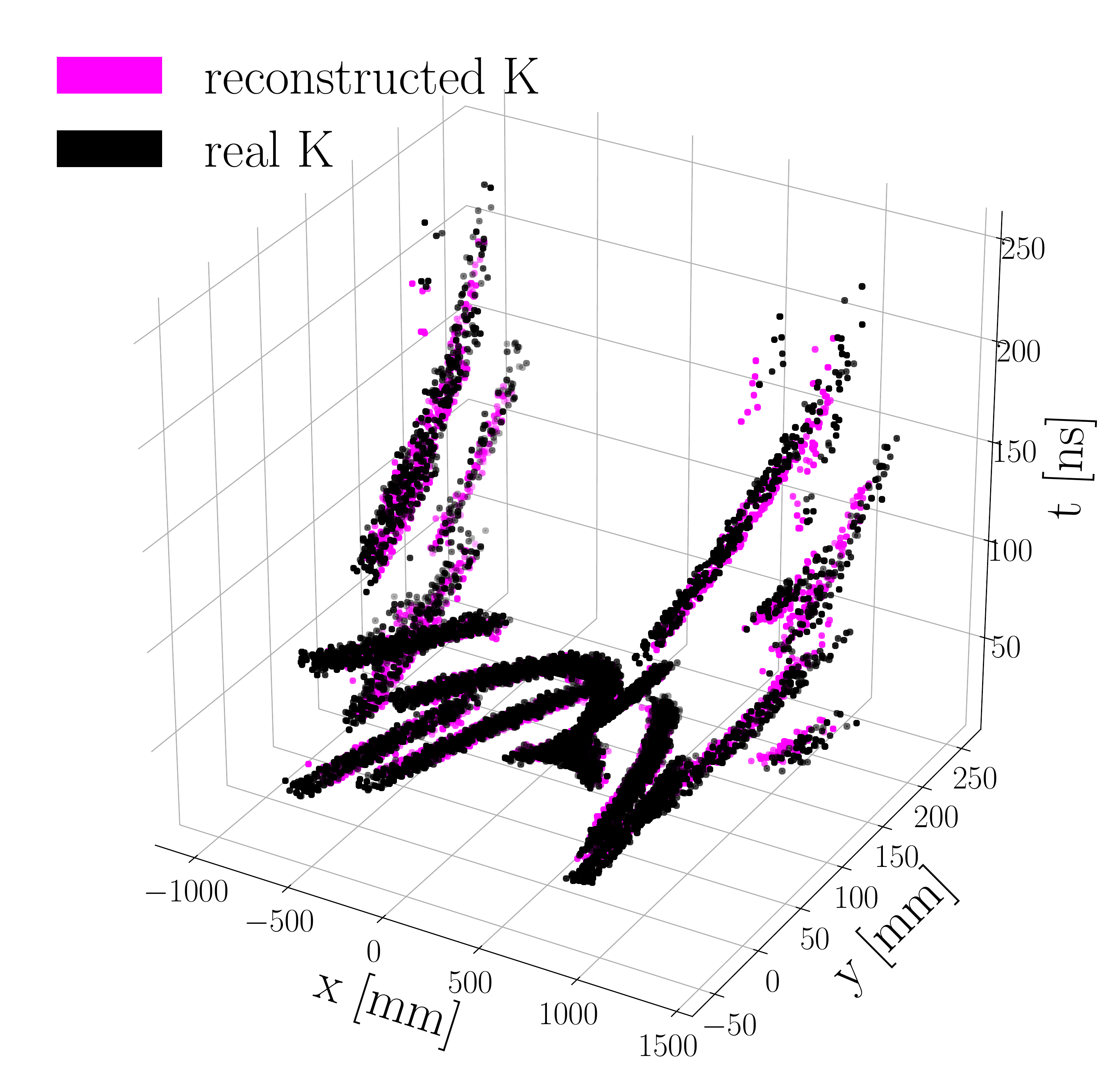}

    \caption{Example of hit points reconstructed by DeepRICH at 4 GeV/c, with an almost perfect overlap between the reconstructed and the injected hits of both pions (left) and kaons (right).
    Image taken from \cite{fanelli2019deeprich}.
    }
    \label{fig:reconstruction}
\end{figure}

The model is trained by minimizing the total loss function which consists of three loss functions, one for reconstruction, one for classification, another one calculated with MMD.  
\\Training samples are prepared in form of discrete hypercubes in the kinematic parameter space ($P,\theta, \phi, X, Y$) of the charged particle. 
The hyperparameters of DeepRICH have been optimized with a Bayesian optimizer.
An example of hit patterns reconstructed by DeepRICH after the tuning of the hyperparameters is shown in Fig. \ref{fig:reconstruction}. 

\begin{figure}[H]
    \centering
    \includegraphics[width=.4\textwidth]{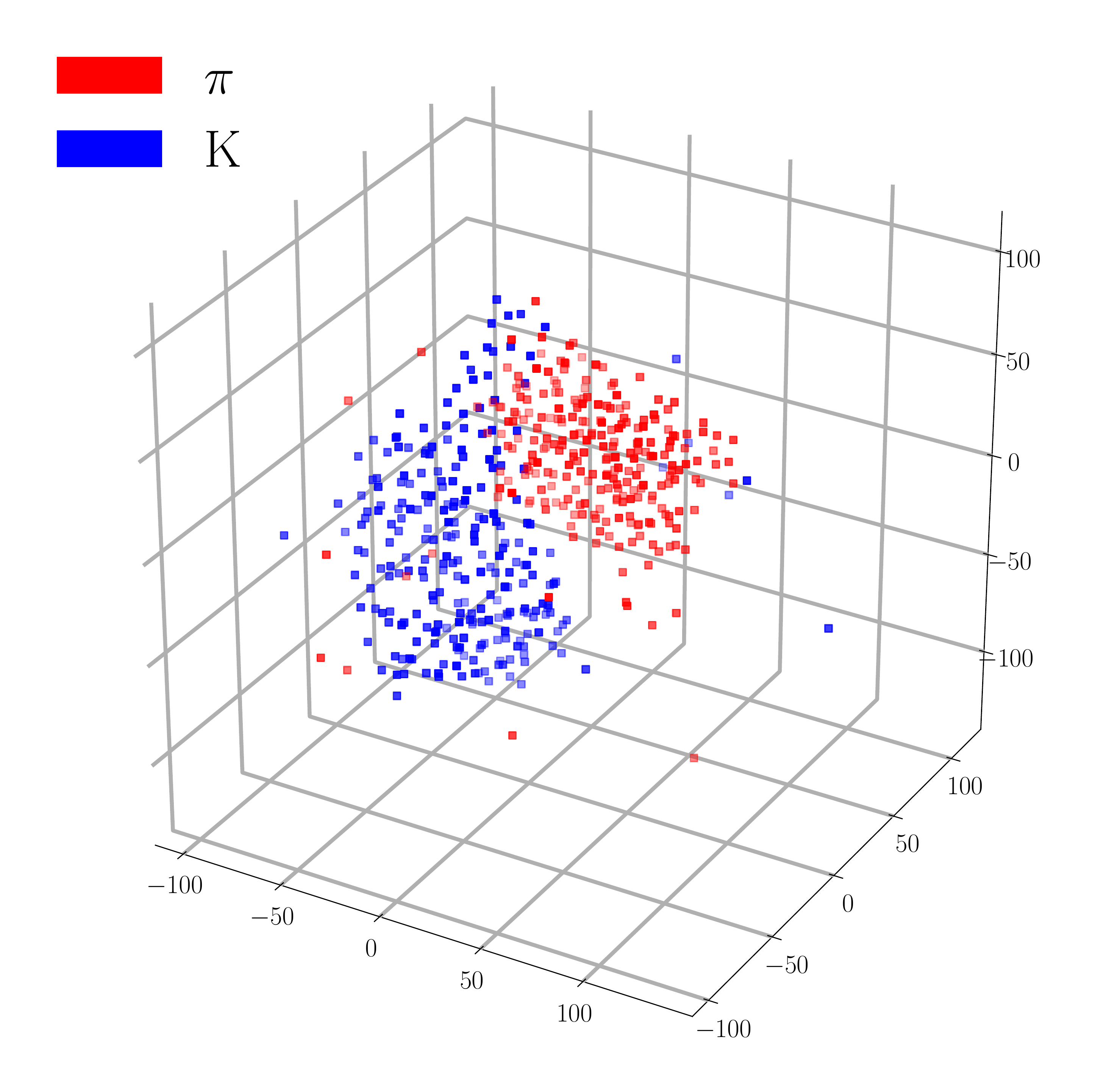}
    \includegraphics[width=.4\textwidth]{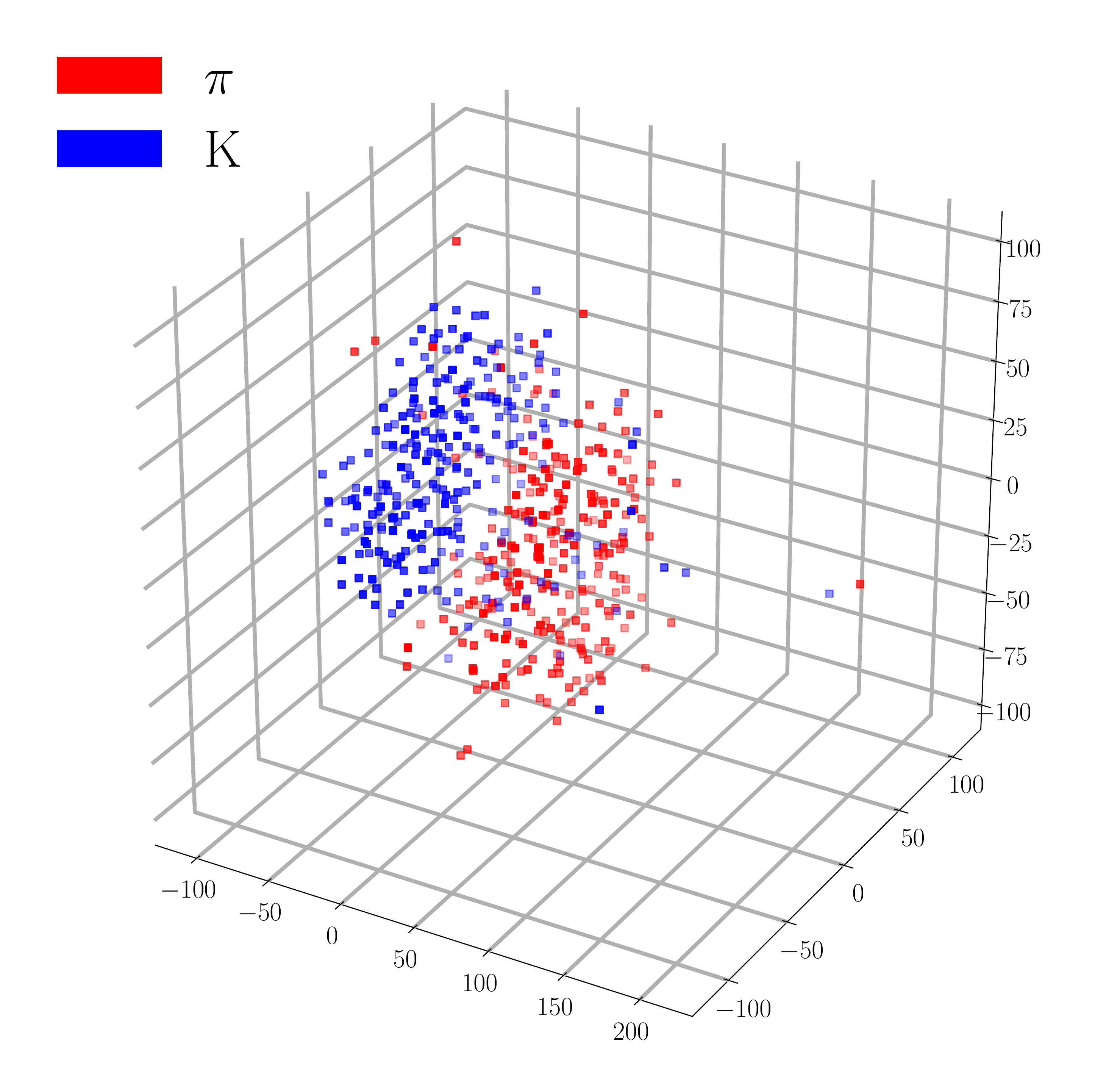}
    \caption{Example of features extracted by the CNN module from $\pi$'s and $K$'s at 4 GeV/c (left) and 5 GeV/c (right).
    These features are then used to classify the particle with a MLP. The plot shows a better separation between $\pi$/$K$ at 4 GeV/c, which means that the network has good distinguishing power. As expected the points become less separated at larger momentum. The 3D visualization is obtained with t-SNE~\cite{maaten2008visualizing}.
    Image taken from \cite{fanelli2019deeprich}.
    }
    \label{fig:pattern1}
\end{figure}

The capability of distinguishing $\pi$s from $K$s and effectively doing PID depends on the features and the causal relations learnt in the space of the latent variables. 
A 3D visualization in the space of the latent variables is shown in Fig. \ref{fig:pattern1}, where dimensionality reduction methods (\textit{viz.}, t-SNE \cite{maaten2008visualizing}) are used to provide a representation of the 20-dimensional space of the latent variables. 
As expected, the larger the momentum the lower is the $\pi/K$ separation.

A thorough comparison with the simulation/reconstruction package FastDIRC is discussed in \cite{fanelli2019deeprich}. 
DeepRICH has the advantage to bypass low-level details needed to build a likelihood and allows to achieve a sensitive improvement in computation time at potentially the same reconstruction performance of other established reconstruction algorithms.

In fact preliminary results show high reconstruction efficiency combined to fast inference time: in particular the time reconstruction of $\mathcal{O}(ms)$ per batch of particles makes DeepRICH potentially faster than established reconstruction methods available at present \cite{fanelli2019deeprich}.

\section{Summary}\label{sec:summary}

AI is entering transversely across different research areas improving the science output. 
In particular we are witnessing a growing number of applications in particle and nuclear physics, beginning with high-level physics analysis and followed by the development of reconstruction algorithms in more recent years.
This paper describes novel directions regarding AI-based applications for imaging Cherenkov detectors. 

Bayesian optimization is currently being used for detector calibration and design optimization. 
The misalignment offsets affecting the instrumentation can be directly inferred from real data with BO, and a real-world example has been described for the \GlueX \ DIRC detector. 
BO can also be used to improve single or multi-detector designs of future experiments like EIC.  
Recent advances in the field show that this can be done in a highly parallelized and automated way.  
With regard to design, real-world costs of the components could be included in the optimization process.
Furthermore, recent optimization packages have been developed to improve the scalability of BO with the number of observations. 

Deep Learning applications have been recently explored to provide fast and accurate reconstruction of charged particles for the DIRC detector. 
The fast reconstruction time in particular, makes a DL-based algorithm suitable for near real-time applications (\textit{e.g.} calibration). 
Another important feature is related to the nature of generative models like VAE, which suggests a tempting scenario of generalizing these new algorithms to fast generation of events once the behavior in the latent space is learnt. 

Finally another suggestive application could be training these algorithms using pure samples of identified particles from real data, allowing to deeply learn the response of the Cherenkov detectors.


\acknowledgments

This material is based upon work supported by the U.S. Department of Energy, Office of Science, Office of Nuclear Physics under contract DE-FG02-94ER40818.

\bibliographystyle{JHEP.bst}
\bibliography{mybibfile}

\end{document}